\documentclass[12pt]{iopart}

\usepackage{subfigure}
\usepackage{epsfig}

\begin{document}

\title[Strange particles in $p+p$ at $\sqrt{\mathrm{s}}$= 200 GeV]{Strange particle production in $p+p$ collisions at $\sqrt{s}$= 200 GeV}

\author{Mark Heinz  \footnote[1]{Electronic address (mheinz@bnl.gov)}
for the STAR Collaboration\footnote[2]{For the full author list and
acknowledgements, see Appendix "Collaborations" in this volume.}}

\address{Laboratory of High Energy Physics, University of Bern,
CH-3012 Switzerland}

\begin{abstract}
We present measurements of the transverse momentum spectra, yield
and $\langle \mathrm{p_{T}} \rangle$ systematics for
$\mathrm{K^{0}_{S}}$, $\Lambda$ and $\bar{\Lambda}$ in $p+p$
collisions at $\sqrt{s}=200$ GeV. We show a dependence of the
$\langle \mathrm{p_{T}} \rangle$ with event multiplicity and infer
that this is consistent with a mini-jet dominated particle
production mechanism. These observations are compared to available
data from $p+\bar{p}$ experiments as well as to pQCD theoretical
predictions.

\end{abstract}

\section{Introduction}

Particles which contain strange quarks are valuable probes of the
dynamics of $p+p$ collisions as constituent strange quarks are not
present in the initial colliding nuclei. The enhancement of the
strange particle yield from $p+p$ collisions to heavy ion collisions
has been suggested as a possible Quark Gluon Plasma signature
\cite{Raf82}.
\\
We report preliminary results from the 2001/2002 p+p run by the STAR
experiment. The main focus of this paper is on presenting a high
statistics measurement of $\mathrm{K^{0}_{S}}$, $\Lambda$ and
$\bar{\Lambda}$ at $\sqrt{s}=200$ GeV and obtaining the yield and
$\langle \mathrm{p_{T}} \rangle$ for each species. A discussion of
appropriate parameterizations to the particle spectra will ensue. We
will show that the widely used power-law extrapolation in
$p+\bar{p}$ collisions is not sufficient to obtain the best
$\chi^{2}$ results for fits to the strange particle spectra and we
will consider alternatives\cite{UA1}.
\\
Furthermore, we will study the dependency of $\langle \mathrm{p_{T}}
\rangle$ and spectral shape as a function of charged particle event
multiplicity($N_{ch}$), an effect that has been observed previously
in high energy $p+\bar{p}$ collisions. Several authors ascribe this
phenomenon to pQCD mini-jets fragmenting into hadronic final states
\cite{Sjo87,Gyu92}. However, we show that present string models
incorporating 'semi-hard' pQCD processes, such as PYTHIA, do not
describe the STAR data well without significant parameter tuning
\cite{Pythia}.

\section{Analysis}

\subsection{Event selection and Pile-up}
The data were reconstructed using the STAR detector system which is
described in more detail elsewhere \cite{STAR1,STAR2}. The main
tracking detector used in this analysis is the Time Projection
Chamber (TPC) covering the full acceptance in azimuth and a large
pseudo-rapidity coverage ($\mid \eta \mid < 1.5$).
\\
A total of 14 million non-singly diffractive (NSD) events were
triggered with the STAR beam-beam counters (BBC) requiring two
coincident charged tracks at forward rapidity ($3.5 < \mid{\eta}\mid
< 5.0$). Due to the particulary low track multiplicity environment
in p+p collisions a special vertex finder was developed. A
Monte-Carlo study showed that only 76\% of primary vertices are
found correctly; from the remainder, 14\% are lost and 10\% are
badly reconstructed. A total of 11 million events passed the event
level selection criteria requiring a valid primary vertex within
100cm along the beam-line from the center of the TPC detector. An
additional multiplicity dependant vertex efficiency factor
$\epsilon_{vtx}(N_{ch})$ must therefore be applied to the particle
spectra.
\\
The high luminosity of the RHIC proton beams has the unfortunate
effect of producing several collisions during the read-out time of
the TPC, a condition known as ``pile-up". These pile-up events
cannot be identified separately, however the triggered vertex can be
identified by allowing only tracks that match to a very fast
scintillating Central Trigger Barrel(CTB) detector to be used in the
vertex reconstruction.

\subsection{Particle reconstruction}
The strange particles were identified from their weak decay to
charged daughter particles via topological identification algorithm.
The following decay channels were analyzed: $\Lambda \rightarrow p +
\pi^{-}$(b.r. 63.9\%) ,$\bar{\Lambda} \rightarrow \bar{p} +
\pi^{+}$(b.r. 63.9\%), $\mathrm{K^{0}_{S}} \rightarrow \pi^{+} +
\pi^{-}$ (b.r. 68.6\%).
\\
Particle identification of daughters was achieved by requiring the
dE/dx to fall within the 3$\sigma$-bands of the theoretical
Bethe-Bloch parameterizations. Further background in the invariant
mass was removed by applying topological cuts to the decay geometry,
 \textit{i.e.} requiring the parent to originate from within 2cm of the primary
vertex and have a minimum decay length of 2cm. These cuts also
ensure that the particle decay is matched to the correct vertex with
a high probability.
\\
Corrections for particle reconstruction efficiency were obtained
through a Monte-Carlo based method of embedding simulated particle
decays into real events and comparing the number of simulated and
reconstructed particles in each $p_{T}$-bin. The $\Lambda$ and
$\bar{\Lambda}$ spectra were corrected for feed-down from $\Xi$
($\Xi^{-} \rightarrow \Lambda + \pi^{-}$). Finally, the corrected
transverse momentum spectra have to be multiplied by the
multiplicity dependant primary vertex efficiency
$\epsilon_{vtx}(N_{ch})$ to account for yield losses.

\section{Transverse Momentum Spectra}

\begin{figure}[h]
\begin{center}
\epsfig{figure=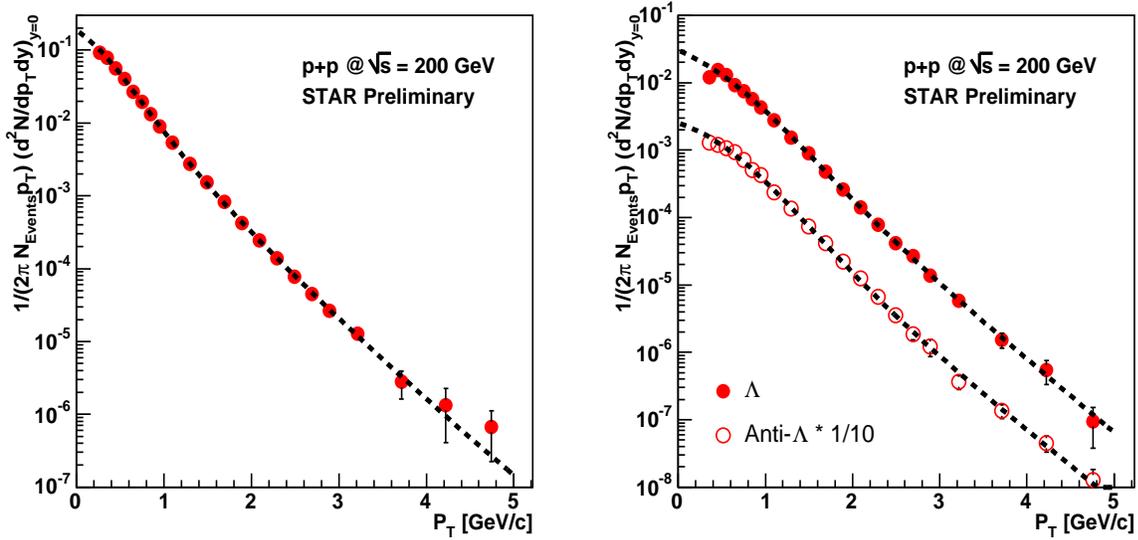, height=8cm, width=16cm}
\caption{Minimum-bias, non-feeddown corrected spectra for
$\mathrm{K^{0}_{S}}$ (left) and $\Lambda$ (right) with ($\mid{y}\mid
\leq 0.5$) from $p+p$ at $\sqrt{s} = 200 $GeV. The errors in the
plot are statistical only. The fit is composite and is described in
the text.} \label{fig:AllSpec}
\end{center}
\end{figure}

Non-feeddown corrected spectra for $\mathrm{K^{0}_{S}}$, $\Lambda$
and $\bar{\Lambda}$ are shown in Figure \ref{fig:AllSpec}. The
particle acceptance at mid-rapidity ($\mid y \mid < 0.5$) in the TPC
starts at transverse momentum 0.2 GeV/c for $\mathrm{K^{0}_{S}}$ and
0.3 GeV/c for $\Lambda$ and $\bar{\Lambda}$. The spectra are
parameterized to allow extrapolation to low pt and to extract
$\langle \mathrm{p_{T}} \rangle$ and yield ($dN/dy$). In contrast to
previous $p+\bar{p}$ experiments \cite{UA1}, which used either a
single exponential function in transverse mass or power-law
functions, we found that a combination of these functions is more
effective in fitting the STAR data. The $\chi^{2}$ results from
different fit-functions are presented in Table \ref{tab:fits1}.
Composite fits, represented in equation 1 (for $\mathrm{K^{0}_{S}}$)
and equation 2 (for $\Lambda$ and $\bar{\Lambda}$) yield the lowest
$\chi^{2}$ and were used to extract the final values for $\langle
\mathrm{p_{T}} \rangle$ and yield as shown in Table
\ref{tab:Results1}.

\begin{table}[h]
\begin{center}
\begin{tabular}{|c|c|c|c|c|c|c|c|}
\hline Particle & $m_{t}$-exponential & power-law & composite equation \\
\hline                      & $\chi^{2}/ndf$   & $\chi^{2}/ndf$   & $\chi^{2}/ndf$ \\
\hline $\Lambda$            & 3.2  & 6.0 & 1.0 \\
\hline $\bar{\Lambda}$      & 4.7  & 4.1 & 1.4 \\
\hline $\mathrm{K^{0}_{S}}$ & 15.9 & 1.6 & 0.7 \\
\hline
\end{tabular}
\caption{A summary of $\chi^{2}$ values for different fit-functions
to the particle spectra } \label{tab:fits1}
\end{center}
\end{table}

\begin{equation}
\frac{1}{2\pi p_{T}}\frac{d^{2}N}{dy dp_{T}} = Ce^{\frac{-m_{T}}{T}}
+ D(1+\frac{p_{T}}{p_{0}})^{-n} \label{Eq:ExpandPower}
\end{equation}

\begin{equation}
\frac{1}{2\pi p_{T}}\frac{d^{2}N}{dy dp_{T}} = Ae^{\frac{-m_{T}}{T}}
+ Be^{\frac{-p_{T}}{T}} \label{Eq:ExpmtandExppt}
\end{equation}

The systematic errors are dominated by the different possible fit
parameterizations and amount to 10\% for $\langle \mathrm{p_{T}}
\rangle$ and 15\% on the yield. The remaining systematical
uncertainties are from the background and efficiency corrections.
Table \ref{tab:ResultsUA5} compares the STAR results to the
measurements by UA5 \cite{UA5}. The UA5 measurements were made over
a larger rapidity interval ($\mid y \mid < 2$) and have been scaled
to match the STAR acceptance by the use of a PYTHIA simulation. The
$\bar{\Lambda}$/$\Lambda$ ratio is 0.88$\pm$0.09 and is flat as a
function of $\mathrm{p_{T}}$, consistent with a nearly net-baryon
free environment at mid-rapidity.

\begin{table}[ht]
\begin{center}
\begin{tabular}{|c|c|c|c|c|c|c|c|}
\hline Particle & extrapolated yield (dN/dy) & feed-down corr. yield (dN/dy) & $\langle \mathrm{p_{T}}\rangle$ (GeV/c)\\
\hline $\Lambda$            & 0.044 $\pm$0.003 & 0.034 $\pm$0.004 & 0.76 $\pm$ 0.01 \\
\hline $\bar{\Lambda}$      & 0.042 $\pm$0.003 & 0.032 $\pm$0.004 & 0.75 $\pm$ 0.01 \\
\hline $\mathrm{K^{0}_{S}}$ & 0.123 $\pm$0.006 &                  & 0.60 $\pm$ 0.01 \\
\hline
\end{tabular}
\caption{A summary of yield and $\langle \mathrm{p_{T}} \rangle$ for
measured strange particles. The errors quoted are statistical only.
} \label{tab:Results1}
\end{center}
\end{table}

\begin{table}[ht]
\begin{center}
\begin{tabular}{|c|c|c|c|c|c|}
\hline Particle  &STAR dN/dy &UA5 dN/dy &UA5 dN/dy &UA5 $\langle \mathrm{p_{T}} \rangle$ \\
&$\mid$y$\mid <$ 0.5&$\mid$y$\mid <$ 2.0 &$\mid$y$\mid <$ 0.5& $\mid$y$\mid <$ 2.0 \\
\hline $\Lambda + \bar{\Lambda}$ & 0.066 $\pm$0.006 & 0.27 $\pm$0.07
&0.076 $\pm$0.02 & 0.8+0.2,-0.14 \\
\hline $\frac{\Lambda + \bar{\Lambda}}{2\mathrm{K^{0}_{S}}}$ &
0.27 $\pm$0.05 & 0.31 $\pm$0.09& 0.31 $\pm$0.09 & 0.53+0.08,-0.06\\
\hline
\end{tabular}
\caption {A summary of yield and $\langle \mathrm{p_{T}} \rangle$
for and $\Lambda$, $\bar{\Lambda}$ (feed down corrected) and
$\mathrm{K^{0}_{S}}$ measured by the UA5 \cite{UA5}. UA5 yields have
been scaled to the STAR measured rapidity interval using a Pythia
simulation} \label{tab:ResultsUA5}
\end{center}
\end{table}

\section{Comparison to PYTHIA}

\begin{figure}[ht]
\begin{center}
\epsfig{figure=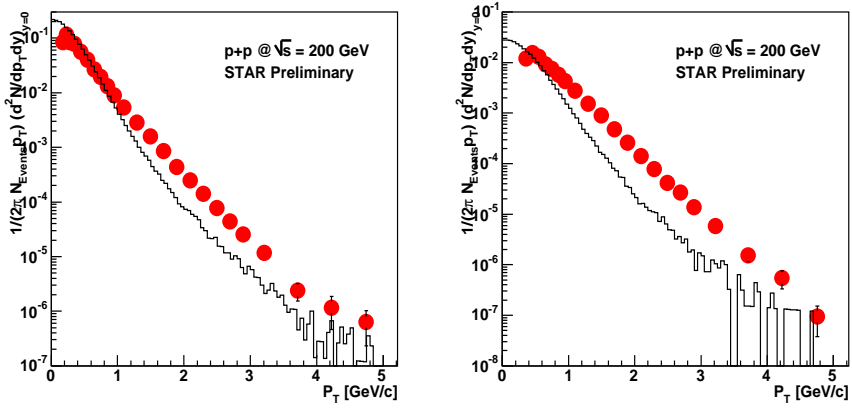, height=6cm, width=12cm}
\caption{$\mathrm{K^{0}_{S}}$ (left) and $\Lambda$ (right) particle
spectra (circles) compared to PYTHIA (line) (ver 6.22, MSEL1)}
\label{fig:SpectraPythia}
\end{center}
\end{figure}

Figure \ref{fig:SpectraPythia} shows the comparison of the STAR
spectra to the spectra from the default setting of PYTHIA event
generator (version 6.22, MSEL1) which includes hard processes and a
modelling of soft processes, through a pQCD model.

Even at $p_{T} \geq$ 1.5, a regime were pQCD should give a better
description, the differences between the model and the data are
considerably large. Further studies are planned in order to
optimally tune the model to match the STAR strangeness data.

\section{$\langle \mathrm{p_{T}} \rangle$ vs event multiplicity}

The minimum bias event sample was split into 6 event classes
according to mean charged particle multiplicity per unit $\eta$
($\langle dN_{ch}/d\eta \rangle$). A similar analysis has been
performed in previous $p+\bar{p}$ experiments at different collision
energies \cite{UA1, E735}. High multiplicity events are particularly
interesting as they are likely dominated by contributions from large
momentum transfer parton-parton collisions.

\begin{figure}[ht]
\begin{center}
\epsfig{figure=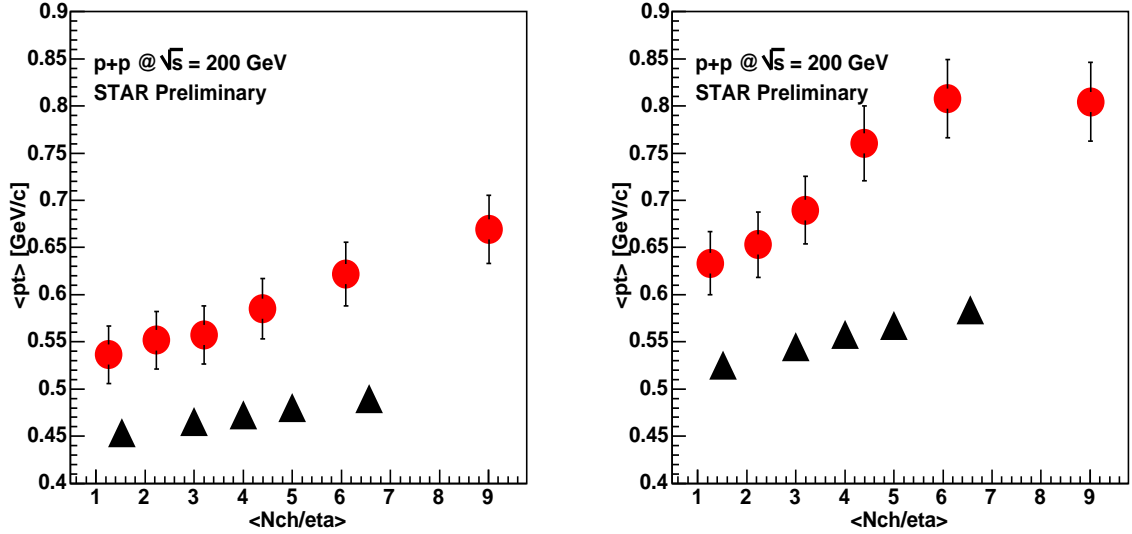, height=8cm, width=16cm}
\caption{$\langle \mathrm{p_{T}} \rangle$ vs. $N_{ch}$ for
$\mathrm{K^{0}_{S}}$ (left) and $\Lambda$ (right) compared to PYTHIA
(triangles) (ver 6.22, MSEL1). The errors on data are both
statistical and systematic.} \label{fig:MeanPtPythia}
\end{center}
\end{figure}

Figure \ref{fig:MeanPtPythia} presents a comparison of the $\langle
\mathrm{p_{T}} \rangle$ vs $\langle dN_{ch}/d\eta \rangle$ for
$\Lambda$ and $\mathrm{K^{0}_{S}}$ with the corresponding PYTHIA
calculations. A rise in $\langle \mathrm{p_{T}} \rangle$ with
increasing $N_{ch}$ is observed and is stronger for the heavier
particle ($\Lambda$). Several authors have attributed these
phenomena to the increased number of fragmenting mini-jets in the
high-multiplicity events \cite{Gai85,Pan86}. In fact, the HIJING
model even makes a prediction for the expected number of mini-jets
contributing to each multiplicity \cite{Gyu92_2}.

It is also interesting that PYTHIA under-predicts the magnitude and
correlation strength of $\langle \mathrm{p_{T}} \rangle$ with
multiplicity. Similar comparisons for inclusive charged particles
$\langle \mathrm{p_{T}} \rangle$ have been recently published by the
CDF Collaboration and demonstrate similar discrepancies
\cite{CDF02}.

\section{$\langle \mathrm{p_{T}} \rangle$ vs particle mass}

\begin{figure}[ht]
\begin{center}
\epsfig{figure=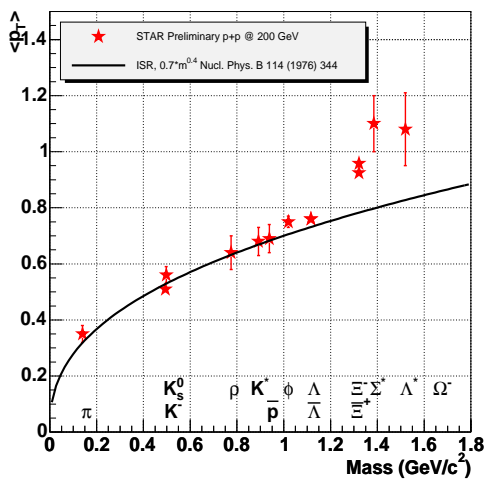, height=7cm, width=7cm}
\caption{$\langle \mathrm{p_{T}} \rangle$ vs particle mass, where
the black curve represents the ISR parametrization from $\pi$, K ,p
at $\sqrt{s}$ = 25 GeV.} \label{fig:MeanPtMass}
\end{center}
\end{figure}

In Figure \ref{fig:MeanPtMass} the $\langle \mathrm{p_{T}} \rangle$
is shown for all particle species measured in $p+p$ collisions in
STAR as a function of particle mass. The black line is an empirical
parameterization to ISR data for $\pi$,K,p only, at $\sqrt{s}$ = 25
GeV. There is very little energy dependance for the lower mass
particles between the ISR and RHIC energies, whereas the higher mass
particles show a clear breaking from the parameterization. The
dependance of $\langle \mathrm{p_{T}} \rangle$ on particle mass has
been explained by mini-jet production in $p+p$ and $p+\bar{p}$
collisions and a stronger contribution for higher mass particles is
expected \cite{Gyu92}.

\section{Summary}

The STAR experiment has made the first high statistics measurement
of mid-rapidity $\mathrm{K^{0}_{S}}$, $\Lambda$ and $\bar{\Lambda}$
in $p+p$ collisions at $\sqrt{s}$ = 200 GeV. The results agree with
those made by the UA5 collaboration for $p+\bar{p}$ at $\sqrt{s}$ =
200 GeV. The ratio of $\bar{\Lambda}$/$\Lambda$ suggests a small net
baryon number at mid-rapidity. We have show that the shape of the
particle spectra is best described with a two-component fit function
accounting for both the soft and hard parts of the spectra.
Furthermore, we have undertaken studies to understand the change in
shape and $\langle \mathrm{p_{T}} \rangle$ of the spectra with
increased event multiplicity. We compared our results to pQCD
inspired models which infer mini-jets and string fragmentation. We
have shown that the current default version of PYTHIA is not well
tuned for strange particles at this collision energy.

\end{document}